\documentclass[
reprint,
superscriptaddress,
amsmath,amssymb,
aps,
prl,
floatfix,
]{revtex4-1}

\usepackage{array}
\usepackage{hhline}
\usepackage{natbib}
\usepackage{color}
\usepackage{graphicx}% Include figure files
\usepackage{dcolumn}% Align table columns on decimal point
\usepackage{multirow}
\usepackage{bm}% bold math
\usepackage{textcomp}
\usepackage{xspace}
\usepackage{hyperref}% add hypertext capabilities
\newcommand{\cs}{$^{125}$Cs\xspace}
\newcommand{\xe}{$^{124}$Xe\xspace}
\newcommand{\xeq}{$^{124}$Xe$^{54+}$\xspace}
\newcommand{\xepg}{$^{124}$Xe(p,$\gamma$)\xspace}
\newcommand{\xepgcs}{$^{124}$Xe(p,$\gamma$)$^{125}$Cs\xspace}
\newcommand{\pg}{(p,$\gamma$)\xspace}
\newcommand{\ag}{($\alpha$,$\gamma$)\xspace}
\newcommand{\tdeg}{\textdegree\xspace}

\newcommand{\hide}[1]{}

\newcommand{\amev}{AMeV\xspace}

\newcolumntype{d}{D{+}{\,\pm\,}{3,3}}
\newcolumntype{x}[1]{>{\centering\arraybackslash\hspace{0pt}}p{#1}}

\newcommand{\gsi}{\affiliation{GSI Helmholtzzentrum f\"ur Schwerionenforschung GmbH, Darmstadt, Germany}}
\newcommand{\guf}{\affiliation{Goethe Universit\"at, Frankfurt am Main, Germany}}
\newcommand{\imp}{\affiliation{Institute of Modern Physics, Chinese Academy of Sciences, Lanzhou, China}}
\newcommand{\edin}{\affiliation{University of Edinburgh, Edinburgh, United Kingdom}}
\newcommand{\atomki}{\affiliation{Institute for Nuclear Research (MTA Atomki), Debrecen, Hungary}}
\newcommand{\cenbg}{\affiliation{CENBG, CNRS-IN2P3, Gradignan, France}}
\newcommand{\jlu}{\affiliation{Justus-Liebig Universit\"at, Gie{\ss}en, Germany}}
\newcommand{\anu}{\affiliation{Australian National University´, Canberra, Australia}}
\newcommand{\hij}{\affiliation{Helmholtz-Insitut Jena, Jena, Germany}}
\newcommand{\spbu}{\affiliation{St. Petersburg State University, St. Petersburg, Russia}}
\newcommand{\bau}{\affiliation{Al-Balqa Applied University, Salt, Jordan}}
\newcommand{\basel}{\affiliation{Department of Physics, University of Basel, Switzerland}}
\newcommand{\herts}{\affiliation{Centre for Astrophysics Research, University of Hertfordshire, Hatfield, United Kingdom}}
\newcommand{\ptb}{\affiliation{Physikalisch-Technische Bundesanstalt, Braunschweig, Germany}}
\newcommand{\tud}{\affiliation{Technische Universit\"at Darmstadt, Darmstadt, Germany}}
\newcommand{\mpi}{\affiliation{Max-Planck-Institut f\"ur Kernphysik (MPIK), Heidelberg, Germany}}
\newcommand{\tub}{\affiliation{Technische Universit\"at Braunschweig, Braunschweig, Germany}}
\begin{document}
\preprint{APS/123-QED}

\title{Approaching the Gamow window with stored ions:\\ Direct measurement of \xepg in the ESR storage ring}
\author{\mbox{J. Glorius}}
\email{j.glorius@gsi.de}
\gsi
\author{\mbox{C. Langer}}
\guf
\author{\mbox{Z. Slavkovsk\'a}}
\guf
\author{\mbox{L. Bott}}
\guf
\author{\mbox{C. Brandau}}
\gsi\jlu
\author{\mbox{B. Br\"uckner}}
\guf
\author{\mbox{K. Blaum}}
\mpi
\author{\mbox{X. Chen}}
\imp
\author{\mbox{S. Dababneh}}
\bau
\author{\mbox{T. Davinson}}
\edin
\author{\mbox{P. Erbacher}}
\guf
\author{\mbox{S. Fiebiger}}
\guf
\author{\mbox{T. Ga{\ss}ner}}
\gsi
\author{\mbox{K. G\"obel}}
\guf
\author{\mbox{M. Groothuis}}
\guf
\author{\mbox{A. Gumberidze}}
\gsi
\author{\mbox{G. Gy\"urky}}
\atomki
\author{\mbox{M. Heil}}
\gsi
\author{\mbox{R. Hess}}
\gsi
\author{\mbox{R. Hensch}}
\guf
\author{\mbox{P. Hillmann}}
\guf
\author{\mbox{P.-M. Hillenbrand}}
\gsi
\author{\mbox{O. Hinrichs}}
\guf
\author{\mbox{B. Jurado}}
\cenbg
\author{\mbox{T. Kausch}}
\guf
\author{\mbox{A. Khodaparast}}
\gsi\guf
\author{\mbox{T. Kisselbach}}
\guf
\author{\mbox{N. Klapper}}
\guf
\author{\mbox{C. Kozhuharov}}
\gsi
\author{\mbox{D. Kurtulgil}}
\guf
\author{\mbox{G. Lane}}
\anu
\author{\mbox{C. Lederer-Woods}}
\edin
\author{\mbox{M. Lestinsky}}
\gsi
\author{\mbox{S. Litvinov}}
\gsi
\author{\mbox{Yu. A. Litvinov}}
\gsi
\author{\mbox{B. L\"oher}}
\tud
\gsi
\author{\mbox{F. Nolden}}
\gsi
\author{\mbox{N. Petridis}}
\gsi
\author{\mbox{U. Popp}}
\gsi
\author{\mbox{T. Rauscher}}
\basel
\herts
\author{\mbox{M. Reed}}
\anu
\author{\mbox{R. Reifarth}}
\guf
\author{\mbox{M. S. Sanjari}}
\gsi
\author{\mbox{D. Savran}}
\gsi
\author{\mbox{H. Simon}}
\gsi
\author{\mbox{U. Spillmann}}
\gsi
\author{\mbox{M. Steck}}
\gsi
\author{\mbox{T. St\"ohlker}}
\gsi\hij
\author{\mbox{J. Stumm}}
\guf
\author{\mbox{A. Surzhykov}}
\ptb
\tub
\author{\mbox{T. Sz\"ucs}}
\atomki
\author{\mbox{T. T. Nguyen}}
\guf
\author{\mbox{A. Taremi Zadeh}}
\guf
\author{\mbox{B. Thomas}}
\guf
\author{\mbox{S. Yu. Torilov}}
\spbu
\author{\mbox{H. T\"ornqvist}}
\gsi\tud
\author{\mbox{M. Tr\"ager}}
\gsi
\author{\mbox{C. Trageser}}
\gsi\jlu
\author{\mbox{S. Trotsenko}}
\gsi
\author{\mbox{L. Varga}}
\gsi
\author{\mbox{M. Volknandt}}
\guf
\author{\mbox{H. Weick}}
\gsi
\author{\mbox{M. Weigand}}
\guf
\author{\mbox{C. Wolf}}
\guf
\author{\mbox{P. J. Woods}}
\edin
\author{\mbox{Y. M. Xing}}
\imp\gsi
\date{\today}

\begin{abstract}
We report the first measurement of low-energy proton-capture cross sections of \xe in a 
heavy-ion storage ring. \xeq ions of five different beam energies between 5.5 \amev and 8 \amev were stored to collide with a 
windowless hydrogen target. 
The \cs reaction products were directly detected.
The interaction energies are located on the high energy tail of the Gamow window for hot, explosive scenarios such as supernovae and X-ray 
binaries. The results serve as an important test of predicted astrophysical reaction rates in this mass range. Good agreement in the 
prediction of the astrophysically important proton width at low energy is found, with only a 30\% difference between 
measurement and theory. Larger deviations are found above the neutron emission threshold, where also neutron- and $\gamma$-widths 
significantly impact the cross sections. 
The newly established experimental method is a very powerful tool to investigate nuclear reactions on rare ion beams at low center-of-mass energies.  
\end{abstract}


\maketitle

Charged-particle induced reactions like \pg and \ag and their reverse reactions play a central role in the quantitative description of explosive scenarios 
like supernovae \cite{travaglio2018} or X-ray binaries \cite{schatz2006}, where temperatures above 1~GK can be reached. The energy 
interval in which the reactions most likely occur under astrophysical conditions is called the Gamow window 
\cite{rolfs1988,rauscher2010}. Experimentalists usually face two major challenges when approaching the Gamow window: firstly, the relatively low center-of-mass energies of only 
a few MeV or less, and secondly, the rapid decrease of cross sections with energy. The high stopping power connected to low-energy beams typically limits the amount of target material, and thus the achievable luminosity. A measurement of small cross sections, on the contrary, requires high luminosities. 

The description of charged-particle processes in explosive nucleosynthesis -- e.g., the $\gamma$ process occurring in core-collapse and thermonuclear supernovae \cite{rauscher2013,rauscher2016,nishimura2018} and the \textit{rp} process on the surface of mass-accreting neutron stars \cite{schatz1998} -- requires large reaction networks including very short-lived nuclei. Experimental data are extremely scarce \cite{kadonis2014}, especially in the mass region $A > 70$, and the modelling relies on calculated cross sections. It is therefore essential to test the theory and its central input parameters. In this Letter we report the first study of the \xepgcs reaction. The cross section is measured on the high energy tail of the Gamow peak, which is located between 2.74 and 5.42~MeV at 3.5~GK in the $\gamma$ process \cite{rauscher2010}. While the \xepg reaction serves as a major milestone for improving the experimental technique to reach lower center-of-mass energies, it also provides important constraints on the so far purely theoretically predicted reaction rates used to model the $\gamma$ process.

In the past, it has been demonstrated at different rare ion beam (RIB) facilities by experiments on ions of mass $A < 40$ \cite{delbar1993,dauria2004,akers2013,lotay2016} that inverse kinematics techniques can be successfully applied to study capture reactions on unstable nuclei. The experiment presented in this Letter has been performed at GSI, Darmstadt \cite{gsi,fair}. Here, the combination of the heavy-ion storage ring ESR \cite{franzke1987} and the FRagment Separator (FRS) \cite{geissel1992} can address all of the aforementioned challenges for reaction measurements in the Gamow window even for the heavy ion beams of interest for the $\gamma$ process. Stored low-energy ions orbit the ring at several hundred kHz, repeatedly impinging on the thin internal target. This recycling of the beam boosts the available luminosity by at least five orders of magnitude, compensating for thin targets and limited beam intensities. 

Recently, a pilot experiment at ESR investigating the reaction $^{96}$Ru(p,$\gamma$)$^{97}$Rh demonstrated the feasibility of this approach for proton-capture reactions \cite{mei2015}. However, only a beam energy of 9 \amev could be reached. For a $^{96}$Ru beam impinging on a $^{1}$H target this converts to a center-of-mass energy of 8.976~MeV, still several MeV away from the Gamow window. Going to even lower energies is a challenge for both, accelerator and experiment.

For the current experiment, the \xe-beam was accelerated to about 100 \amev in the UNIversal Linear ACcelerator (UNILAC) and SchwerIonenSynchroton (SIS18), extracted to the ESR transfer beam line, completely stripped off bound electrons and finally injected into the ESR. Once the beam was stored, its momentum spread was brought down to and maintained at $\frac{\Delta p}{p} \approx 10^{-5}$ with the electron cooling system of the ring \cite{litvinov2013}. In the next step, the ions were decelerated to the desired energies of a few \amev. About $10^6$ - $10^7$ \xeq ions at energies as low as 3 \amev can potentially be stored in the ESR. At this point the internal ultra-pure H$_2$ target was switched on reaching densities of about $10^{14}$ atoms/cm$^{2}$ \cite{target}. This corresponds to an energy loss of about 5-10~eV, which is compensated by the electron cooler. The beam passed through the hydrogen target with a revolution frequency of about 250 - 500 kHz, resulting in peak luminosities of about $L = 10^{26} ~\text{cm}^{-2}\text{s}^{-1}$. Measurements at 5 beam energies starting from 8 \amev and reaching down as low as 5.5 \amev were performed to investigate the \xepg reaction in inverse kinematics. Atomic interactions with the atoms of the target and the residual gas limit the storage time of highly charged ions at low energies. With the \xeq beam stored at 7 \amev a beam lifetime of about 2.5~s could be achieved, resulting in a reasonable measurement period of about 12~s before the ring had to be refilled. A single fill cycle of the ESR took about 50 s, implying a duty cycle of about 25\%.

\begin{figure}[t]
\centering\includegraphics[width=\linewidth]{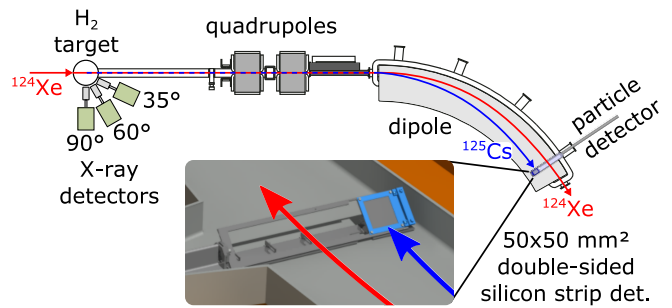}
\caption{The figure shows the experimental setup at the ESR from the gas target to the next dipole magnet. Three Ge X-ray detectors are placed around the interaction region at angles of 35\tdeg, 60\tdeg and 90\tdeg. The DSSSD is positioned in the last quarter of the dipole to intercept the produced \cs ions, which are separated from the circulating \xe beam due to their magnetic rigidity.}\label{fig:setup}
\end{figure}

The \cs products of the \pg reaction are subject to a negligible momentum recoil caused by the emission of the $\gamma$ cascade, see \textit{e.g.} \cite{ruiz2014}. This allows the entire recoil cone to be covered by a single particle detector. In order to separate the reaction products from the stored beam, the detection system is implemented at the end of the first dipole magnet downstream of the target as shown in Fig.\ \ref{fig:setup}. The lower magnetic rigidity results in a separation from the stored beam of about 4~cm, which was also predicted by beam-optical simulations. It should be noted that for an undisturbed detection of the \pg reaction products, it is essential to utilize a fully stripped primary beam. Otherwise, the stored ions which loose an electron at the target would hit the detector at approximately the same position as the \pg products, due to a comparable magnetic rigidity. This would lead to a fatal background contribution, since at low energies the cross section for ionization is much larger than for proton capture.

The main challenges for this experiment were the storage and detection of ions at Gamow window energies. In order to store highly charged ions at energies below 10 \amev ultra-high vacuum (UHV) conditions of about $10^{-11}$ mbar in the entire ring are crucial. Otherwise the atomic interactions of the revolving beam with the residual gas atoms would reduce possible storage times to the sub-second level \cite{grieser2012,stoehlker1998}, which would render reaction studies impossible. These boundary conditions dictate a highly restrictive list of materials that can be brought into the UHV environment. Therefore, the regular particle detection systems at the ESR are operated inside detector pockets, which are separated from the ring vacuum by entrance windows made of 25-100 $\mu$m stainless steel \cite{klepper2003}. As heavy ions of energies below 10 \amev hardly penetrate such windows, the design and implementation of a new in-vacuum detection system was the major step towards the low energies of the Gamow window.

The new in-vacuum setup consists of a Micron Semiconductor Ltd W1-type double-sided silicon strip detector (DSSSD) \cite{dsssd} of 500 $\mu$m thickness, which stops low-energy ions completely. To be compatible to the UHV environment, the wafer is mounted on a ceramic printed-circuit board, equipped with low-outgassing cables and designed for in-situ bakeout at about 125 \tdeg{C}. The DSSSD is able to detect ion energy deposits of several hundred MeV at 100\% efficiency with a spatial resolution of about 3 mm and an energy resolution better than 1\%. The setup has been installed at 53.5\tdeg ~bending angle of the 60\tdeg ~dipole magnet downstream of the target as indicated in Fig.\ \ref{fig:setup}.

To extract absolute \pg cross section values in the analysis, the luminosity in the ring has to be known. It depends on the areal thickness of the target, the beam current and their mutual geometric overlap. For this purpose, the investigation of the \pg cross section is carried out relative to a measurement of the radiative electron-capture process from the H$_2$ target to the K-shell of \xeq (K-REC). High-purity germanium semiconductor detectors surrounding the target at 90\tdeg, 60\tdeg and 35\tdeg with respect to the beam axis were used to detect the X-ray signature of the K-REC. The REC process is one of the dominant processes in ion-atom (ion-electron) collisions and has been studied in details in recent decades \cite{eichler2007}. These studies have demonstrated that all experimental REC results can be well understood within the framework of the relativistic distorted-wave approach. Based on this approach the K-REC differential cross sections can be predicted with an uncertainty $\leq$ 2\%. The main source of this uncertainty arises from the fact that a molecular H$_2$ target is used instead of atomic H \cite{artemyev2010}. 

\begin{figure}[b]
\centering\includegraphics[width=\linewidth]{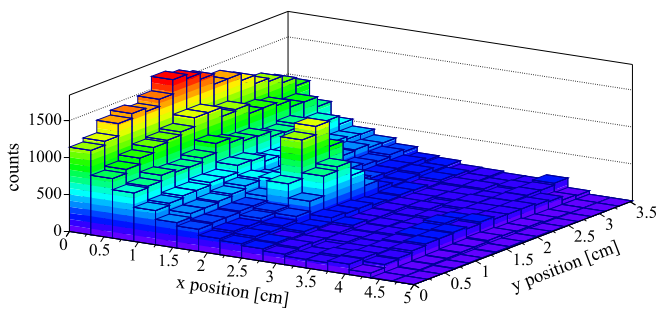}
\caption{The ion hit distribution measured with the DSSSD is shown. On top of a broad background of elastically scattered \xe ions a narrow cluster of \cs ions can be identified as products of the \pg reaction.}\label{fig:dsssd}
\end{figure}

For all beam energies the DSSSD was positioned about 1 cm away from the orbit of the beam to capture the \pg products in the center of the active area. The hit distribution across the surface of the DSSSD at 7 \amev is shown in Fig.\ \ref{fig:dsssd}. The narrow cluster of \cs ions from the \pg reaction in the center of the detector is clearly visible above the broad background of \xe ions from Rutherford elastic scattering off the hydrogen target. 

For ion detection with the DSSSD a coincidence condition between front and back side of the detector in combination with a simple energy threshold at 1/3 of the nominal ion energy has been applied. This leads to a clean ion-hit identification also taking into account inter-strip events, which result in energy sharing between adjacent strips \cite{yorkston1987}. Since no notable losses have been observed, an ion detection efficiency of 100\% is assumed.

The determination of the \pg cross section $\sigma_{\mathrm{(p,}\gamma \mathrm{)}}$ can be described as
\begin{equation}
\sigma_{\mathrm{(p,}\gamma \mathrm{)}} = N_{\mathrm{(p,}\gamma \mathrm{)}}\frac{\epsilon_{\textrm{K}}\Delta \Omega}{N_{\textrm{K}}} \frac{d\sigma_{\textrm{K}}}{d\Omega}~.
\label{eq:pg}
\end{equation}

Here, $N_{\mathrm{(p,}\gamma \mathrm{)}}$ is the number of detected \cs ions, $N_{\textrm{K}}$ denotes the number of K-REC X-rays detected with the efficiency $\epsilon_{\textrm{K}}$ within the solid angle $\Delta \Omega$, and $d\sigma_{\textrm{K}}/d\Omega$ is the K-REC differential cross section.

The extraction of the number of proton-capture events from the 2-dimensional histograms was accomplished by fitting and subtracting the Rutherford background. The shape of this background component was simulated with the Monte Carlo based MOCADI code \cite{iwasa2011}. The simulation took into account the well-known Rutherford scattering kinematics and angular distribution \cite{rutherford1911} as well as all ion-optical elements, such as quadrupole and dipole magnets, which deform and shift the initial distribution. Additionally, background from nuclear channels such as (p,n) and $\mathrm{(p,}\alpha \mathrm{)}$ has been investigated and was found to be insignificant, either due to a clear separation in the dipole field or due to a negligible cross section as predicted by theory. This is confirmed by the Rutherford background fits, which describe the experimental data with $\chi^{2}_\mathrm{reduced}$ values close to 1. The residual ion hits after background subtraction are concentrated in a narrow cluster, which was integrated to obtain the number of \pg reaction products $N_{\mathrm{(p,}\gamma \mathrm{)}}$. 

\begin{figure}[t!]
\centering\includegraphics[width=\linewidth]{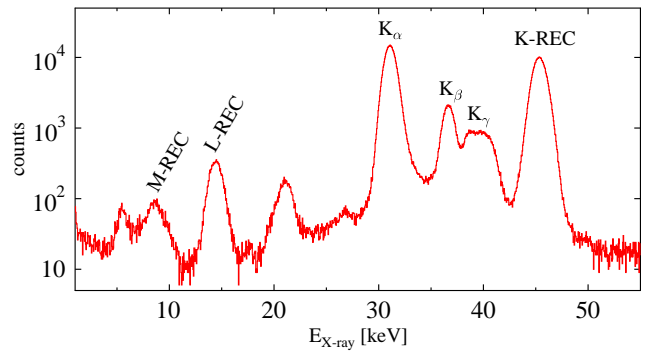}
\caption{Depicted is the spectrum of X-ray radiation recorded by the 90\tdeg detector at the target with a beam energy of 8 \amev. The K-REC peak at 45~keV is used for luminosity normalization. Various other atomic processes are visible through their characteristic lines. For details see text. }\label{fig:xray}
\end{figure}

For normalization the K-REC process is measured by means of X-ray spectroscopy. A typical spectrum taken at 8 \amev is shown in Fig.\ \ref{fig:xray}. In addition to the K-REC signature at about 45~keV, the Lyman series ($K_\alpha, K_\beta, K_\gamma$) is clearly visible, as well as the REC processes involving the L- and M-shell of xenon. Unmarked peaks are due to detector response or could not be identified unambiguously. The X-ray spectra were recorded simultaneously with the \pg spectra in the DSSSD. The number of K-REC counts, $N_{\textrm{K}}$, resulted from integrating the K-REC peak assuming a linear background.

The energy-dependent X-ray detection efficiency was measured using calibrated radioactive sources ($^{133}$Ba,$^{210}$Pb,$^{241}$Am) at exactly the same distance from the detector as the hydrogen target during the experiment, which includes a measurement of the solid angle $\Delta \Omega$. The interpolation to K-REC energies was done by fitting the measured efficiency data. 

The X-ray normalization procedure has been confirmed by an independent luminosity determination making use of the background distribution detected with the DSSSD based on the Rutherford formula \cite{rutherford1911}. Due to uncertainties in detector position and beam-optical parameters, as well as from nuclear contribution to the scattering cross section \cite{nurmela1998}, the uncertainty of this method is on the order of 10-20\%, but the results agree with the X-ray normalization within 10\%.

\begin{table*}[t!]
\caption{Final results for the \xepg cross section and interim results used in Eqs.\ (\ref{eq:pg}) to (\ref{eq:n2}). See text for details.}
\label{tab:values}
\centering
\begin{tabular}{x{1.3cm}x{3.5cm}x{0.1cm}x{2cm}x{2cm}x{2cm}x{2cm}x{1.2cm}x{1.2cm}x{1.2cm}}\cline{1-2}\cline{4-10}\\[-2mm]%\hhline{==~=======}\\[-2mm]
$E_{\textrm{CM}}$ & $\sigma_{\mathrm{(p,}\gamma \mathrm{)}}$ & & \multirow{2}{*}{$N_{\mathrm{(p,}\gamma \mathrm{)}}$} &  \multicolumn{3}{c}{$N_{\textrm{K}}/\epsilon_{\textrm{K}}\Delta \Omega \left[10^6/\textrm{sr}\right]$ } & \multicolumn{3}{c}{$d\sigma_{\textrm{K}}/d\Omega \left[\textrm{barn/sr}\right]$} \\[1mm]
[MeV]& [mbarn] & & & 90\tdeg & 60\tdeg & 35\tdeg & 90\tdeg & 60\tdeg & 35\tdeg  \\[1mm]
\cline{1-2}\cline{4-10}\\[-3mm]
$5.47$ & $14.0 \pm 2.4_{\textrm{stat}} \pm 0.9_{\textrm{syst}}$ & & $\hphantom{1}785 \pm 134$ & $13.51\pm1.06$ & $10.73\pm0.55$ & $4.86\pm0.28$ & $244.5$ & $190.2$ & $86.4$ \\
$5.95$ & $28.0 \pm 2.6_{\textrm{stat}} \pm 1.9_{\textrm{syst}}$ & & $1591 \pm 149$ & $12.54\pm0.98$ & $\hphantom{1}9.93\pm0.51$ & $-$     & $223.8$ & $174.2$ & $79.1$ \\
$6.65$ & $65.8 \pm 4.4_{\textrm{stat}} \pm 4.2_{\textrm{syst}}$ & & $1280 \pm \hphantom{1}85$ & $\hphantom{1}3.91\pm0.31$ & $\hphantom{1}3.03\pm0.15$ & $1.35\pm0.08$   & $199.0$ & $154.9$ & $70.4$ \\
$6.96$ & $97.7 \pm 2.7_{\textrm{stat}} \pm 6.7_{\textrm{syst}}$ & & $5500 \pm 153$ & $10.53\pm0.82$ & $\hphantom{1}8.36\pm0.43$ & $-$     & $189.7$ & $147.7$ & $67.1$ \\
$7.92$ & $43.2 \pm 2.2_{\textrm{stat}} \pm 2.8_{\textrm{syst}}$ & & $2774 \pm 141$ & $10.40\pm0.81$ & $\hphantom{1}8.35\pm0.42$ & $3.74\pm0.22$ & $165.3$ & $128.8$ & $58.6$ \\\cline{1-2}\cline{4-10}
%\hhline{--~-------}
\end{tabular}
\end{table*}

The measurements at 7 \amev and 6 \amev were carried out with two X-ray detectors at 90\tdeg and 60\tdeg. For the runs taken at 5.5 \amev, 6.7 \amev and 8 \amev an additional detector at 35\tdeg was available. For each beam energy the \pg cross sections based on normalization to individual X-ray detectors agreed within the uncertainties. The final results were obtained from the weighted average of the individual normalization factors $F_{\textrm{norm},i}$:
\begin{eqnarray}
F_{\textrm{norm},i} &=& \frac{\epsilon_{\textrm{K},i}\Delta \Omega_i}{N_{\textrm{K},i}} \frac{d\sigma_{\textrm{K}}}{d\Omega}(\theta_i)~,\label{eq:n1}\\
\left<F_{\textrm{norm}}\right> &=& \left. \sum_i \frac{F_{\textrm{norm},i}}{\sigma_i^2} \middle/ \sum_i \frac{1}{\sigma_i^2} \right.
,\label{eq:n2}
\end{eqnarray}
where $\sigma_i$ is the individual uncertainty associated with $F_{\textrm{norm},i}$ not taking into account common uncertainties, like the one connected to the K-REC cross section. 

The final cross section values are listed in Tab.\ \ref{tab:values} for all center-of-mass energies. These energies are determined from the applied voltage at the electron cooler as demonstrated by \cite{ullmann2015}. The error in the voltage measurement leads to an uncertainty for $E_{\textrm{CM}}$ of about 10~keV. The cross-section uncertainties for the two lowest energies are dominated by the statistical component of $N_{\mathrm{(p,}\gamma \mathrm{)}}$, which also includes the uncertainty of the Rutherford fit. At higher beam energies the systematic component becomes equally important, which mainly originates from the X-ray normalization, including uncertainties from the K-REC cross section (2\%), the calibration standards (5\%) and the X-ray efficiency (5\%). It has to be noted that the latter is subject to averaging according to Eq.\ (\ref{eq:n2}). The results of intermediate analysis steps are also provided in Tab.\ \ref{tab:values}. Here, the individual efficiency-corrected K-REC counts per steradian $N_{\textrm{K-REC}}/\epsilon_{\textrm{K-REC}}\Delta \Omega$ as well as the effective theoretical $d\sigma_{\textrm{K}}/d\Omega$ are listed separately for all beam energies and available detector angles. Together with the number of \pg products $N_{\mathrm{(p,}\gamma \mathrm{)}}$ the full dataset for use in Eqs.\ (\ref{eq:pg}) to (\ref{eq:n2}) is available.

\begin{figure}[b]
\centering\includegraphics[width=\linewidth]{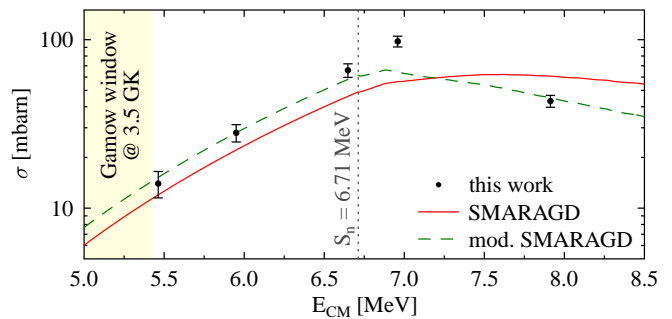}
\caption{Comparison of experimental and predicted cross sections: the red solid line shows the predictions obtained with the same
input as used in the reaction rate libraries for astrophysics; the green dashed line shows a similar calculation but with 
the proton width increased by 30\% and the neutron width increased by a factor of 2.5 (see text for details).\label{fig:xs}}
\end{figure}

Theoretical nuclear cross sections for medium- and heavy-mass nuclei at astrophysical energies are usually calculated within the Hauser-Feshbach (HF) formalism based on the formation of a compound nucleus at high level density \cite{hauserfeshbach,rauscher11}. Formation and decay probabilities of the compound system are quantified in so-called averaged widths. The majority of reaction rates used in large nucleosynthesis networks are based on such HF calculations. We used our new data to test the HF cross section predictions and the underlying physical parameters directly relevant for the $\gamma$ process \cite{rauscher2013}. For this purpose we use the HF code SMARAGD \cite{rauscher11,smaragd}, which relies on a widely used combination of nuclear models for astrophysical rate prediction \cite{rauscher00,cyburt2010}. In detail, the nuclear input to the HF code consists of the microscopic proton+nucleus optical model potential by \cite{jeukenne77} with low-energy modifications by \cite{lejeune80}, the nuclear level density by \cite{rtk} including a parity dependence according to \cite{mocelj}, and the gamma-strength as described in \cite{rauscher00,rtk}.  

Of special astrophysical interest is the energy range below the neutron-emission threshold $S_{n} = 6.71$ MeV \cite{rauscher2010,rauscher2013,nishimura2018}. In this region the proton width is typically the most sensitive input parameter for \pg cross sections \cite{rauscher12}, making it a key ingredient for HF rate predictions in the $\gamma$ process \cite{rauscher2013}. In Fig.\ \ref{fig:xs} the experimental cross section values are compared to results of the SMARAGD code. For the unmodified SMARAGD prediction the deviation of about 30\% at the lower end of the measured energy range is within the expectations, since the underlying models aim at a global description of nuclear properties. However, better agreement can be obtained by locally increasing the proton width for the compound nucleus \cs by 30\% which increases the calculated cross section by about the same amount. This confirms that the usually adopted uncertainty of about a factor of $2$ for global HF rate predictions of \pg and inverse reactions \cite{rauscher2016} holds in this mass region. 

Above the neutron-emission threshold, the cross-section predictions are additionally sensitive to the $\gamma$- and neutron widths. Therefore it is impossible to unambiguously identify the source of the strong deviation between theory and experiment at these energies. Varying all widths shows that it is impossible to simultaneously reproduce the data points at the two highest energies. Either the cross section at the highest measured energy is strongly overpredicted or the data point at the second-highest energy is underpredicted. The latter case is shown in Fig.\ \ref{fig:xs}, which requires a strong increase in the neutron width or alternatively a strong decrease of the $\gamma$ width in the compound nucleus to bring down the cross section to coincide with the data point at the highest energy. Such a strong variation would be typical for a single resonance structure but this would not be expected at the high level density found in the compound nucleus \cs at the populated excitation energies.

In Summary, with the study of \xepgcs presented here, the in-ring method for direct \pg measurement was proven to be applicable for heavy nuclides and to provide measurements in an astrophysically relevant energy range. In combination with the rare ion beam production in the FRS at GSI \cite{geissel1992} this offers new opportunities to significantly improve weakly constrained astrophysical reaction rates used in nuclear reaction networks responsible for the synthesis of the \textit{p} nuclei in explosive stellar scenarios. The present data for \xepg between 5.5 and 8 MeV provide a sensitive test of cross section predictions and especially the prediction of the proton width which is important for the theoretical modelling of astrophysical proton capture and inverse reactions. Although several modifications of theoretical parameters were needed to describe the experimental dataset, we can confirm that the theory provides reliable \pg predictions within the assumed accuracy of about a factor of two. This especially holds for the proton width and the lower part of the measured energy range. 

In the future it is envisioned to extend the proton-capture campaign to radioactive beam studies, addressing key reactions like $^{59}$Cu(p,$\gamma$), which has high impact on the light curve and heavy element production in X-ray burst models \cite{cyburt2016}. The only strong limitation for radioactive beam experiments at the present ESR facility is the half-life of the stored ion; it should be on the order of tens of seconds at least. Moreover, with the in-ring technique it is conceivable to broaden the range of reaction channels that can be studied in inverse kinematics, e.g. ($\alpha$,$\gamma$) or (p,n) reactions would be possible, with only minor modifications to the experimental setup.

Driven by the high scientific potential, there are several initiatives for new storage-ring facilities around the world with a focus on low-energy studies. The storage ring at HIE-ISOLDE project \cite{grieser2012}, for instance, combines a low-energy ring with an ISOL-type RIB facility, while the CRYRING@ESR project \cite{lestinsky2016} represents a low-energy extension of the ESR machine. This work delivers a proof-of-principle for one of the key physics cases connected to such ring projects.

\begin{acknowledgments}
This project has received funding from the European Research Council (ERC) under the European Union's Horizon 2020 research and innovation programme (grant agreement No 682841 ``ASTRUm'').
This work is supported by the Helmholtz International Center for FAIR (HIC for FAIR),
by the Bundesministerium f\"ur Bildung und Forschung (BMBF) (05P15RFFAA, 05P15RGFAA),
by the Science and Technology Facilities Council (STFC) UK (ST/L005824/1, ST/M001652/1, ST/M006085),
by the Helmholtz-CAS Joint Research Group (HCJRG-108), 
and by the Helmholtz-OCPC Postdoctoral Program 2017 (GSI08).
CLW acknowledges support by the European Research Council (grant agreement ERC-2015-StG Nr. 677497 ``DoRES'').
SD gratefully acknowledges the support provided by the Alexander von Humboldt Foundation and the Jordanian Scientific Research Support Fund under grant \#Bas/2/4/2014. 
SYuT acknowledges the support by the DAAD through Mendeleev grant, SPbU(28999675). 
CL and JG thank the ExtreMe Matter Institute EMMI at GSI, Darmstadt, and JINA-CEE (NSF grant No. PHY-1430152) for support in the framework of the EMMI/JINA-CEE Workshop "NARRS" during which this work was discussed.
\end{acknowledgments}


\begin{thebibliography}{44}
\expandafter\ifx\csname natexlab\endcsname\relax\def\natexlab#1{#1}\fi
\expandafter\ifx\csname bibnamefont\endcsname\relax
  \def\bibnamefont#1{#1}\fi
\expandafter\ifx\csname bibfnamefont\endcsname\relax
  \def\bibfnamefont#1{#1}\fi
\expandafter\ifx\csname citenamefont\endcsname\relax
  \def\citenamefont#1{#1}\fi
\expandafter\ifx\csname url\endcsname\relax
  \def\url#1{\texttt{#1}}\fi
\expandafter\ifx\csname urlprefix\endcsname\relax\def\urlprefix{URL }\fi
\providecommand{\bibinfo}[2]{#2}
\providecommand{\eprint}[2][]{\url{#2}}

\bibitem[{\citenamefont{Travaglio et~al.}(2018)\citenamefont{Travaglio,
  Rauscher, Heger, Pignatari, and West}}]{travaglio2018}
\bibinfo{author}{\bibfnamefont{C.}~\bibnamefont{Travaglio}},
  \bibinfo{author}{\bibfnamefont{T.}~\bibnamefont{Rauscher}},
  \bibinfo{author}{\bibfnamefont{A.}~\bibnamefont{Heger}},
  \bibinfo{author}{\bibfnamefont{M.}~\bibnamefont{Pignatari}},
  \bibnamefont{and} \bibinfo{author}{\bibfnamefont{C.}~\bibnamefont{West}},
  \bibinfo{journal}{The Astrophysical Journal} \textbf{\bibinfo{volume}{854}},
  \bibinfo{pages}{18} (\bibinfo{year}{2018}).

\bibitem[{\citenamefont{Schatz and Rehm}(2006)}]{schatz2006}
\bibinfo{author}{\bibfnamefont{H.}~\bibnamefont{Schatz}} \bibnamefont{and}
  \bibinfo{author}{\bibfnamefont{K.}~\bibnamefont{Rehm}},
  \bibinfo{journal}{Nuclear Physics A} \textbf{\bibinfo{volume}{777}},
  \bibinfo{pages}{601} (\bibinfo{year}{2006}).

\bibitem[{\citenamefont{Rolfs and Rodney}(1988)}]{rolfs1988}
\bibinfo{author}{\bibfnamefont{C.}~\bibnamefont{Rolfs}} \bibnamefont{and}
  \bibinfo{author}{\bibfnamefont{W.}~\bibnamefont{Rodney}},
  \emph{\bibinfo{title}{Cauldrons in the {Cosmos}: {Nuclear} {Astrophysics}}}
  (\bibinfo{publisher}{University of Chicago Press}, \bibinfo{year}{1988}).

\bibitem[{\citenamefont{Rauscher}(2010)}]{rauscher2010}
\bibinfo{author}{\bibfnamefont{T.}~\bibnamefont{Rauscher}},
  \bibinfo{journal}{Physical Review C} \textbf{\bibinfo{volume}{81}},
  \bibinfo{pages}{045807} (\bibinfo{year}{2010}).

\bibitem[{\citenamefont{Rauscher et~al.}(2013)\citenamefont{Rauscher, Dauphas,
  Dillmann, Fr\"ohlich, F\"ul\"op, and Gy\"urky}}]{rauscher2013}
\bibinfo{author}{\bibfnamefont{T.}~\bibnamefont{Rauscher}},
  \bibinfo{author}{\bibfnamefont{N.}~\bibnamefont{Dauphas}},
  \bibinfo{author}{\bibfnamefont{I.}~\bibnamefont{Dillmann}},
  \bibinfo{author}{\bibfnamefont{C.}~\bibnamefont{Fr\"ohlich}},
  \bibinfo{author}{\bibfnamefont{Z.}~\bibnamefont{F\"ul\"op}},
  \bibnamefont{and} \bibinfo{author}{\bibfnamefont{G.}~\bibnamefont{Gy\"urky}},
  \bibinfo{journal}{Reports on Progress in Physics}
  \textbf{\bibinfo{volume}{76}}, \bibinfo{pages}{066201}
  (\bibinfo{year}{2013}).

\bibitem[{\citenamefont{Rauscher et~al.}(2016)\citenamefont{Rauscher,
  Nishimura, Hirschi, Cescutti, Murphy, and Heger}}]{rauscher2016}
\bibinfo{author}{\bibfnamefont{T.}~\bibnamefont{Rauscher}},
  \bibinfo{author}{\bibfnamefont{N.}~\bibnamefont{Nishimura}},
  \bibinfo{author}{\bibfnamefont{R.}~\bibnamefont{Hirschi}},
  \bibinfo{author}{\bibfnamefont{G.}~\bibnamefont{Cescutti}},
  \bibinfo{author}{\bibfnamefont{A.~S.} \bibnamefont{Murphy}},
  \bibnamefont{and} \bibinfo{author}{\bibfnamefont{A.}~\bibnamefont{Heger}},
  \bibinfo{journal}{Monthly Notices of the Royal Astronomical Society}
  \textbf{\bibinfo{volume}{463}}, \bibinfo{pages}{4153} (\bibinfo{year}{2016}).

\bibitem[{\citenamefont{Nishimura et~al.}(2018)\citenamefont{Nishimura,
  Rauscher, Hirschi, Murphy, Cescutti, and Travaglio}}]{nishimura2018}
\bibinfo{author}{\bibfnamefont{N.}~\bibnamefont{Nishimura}},
  \bibinfo{author}{\bibfnamefont{T.}~\bibnamefont{Rauscher}},
  \bibinfo{author}{\bibfnamefont{R.}~\bibnamefont{Hirschi}},
  \bibinfo{author}{\bibfnamefont{A.~S.} \bibnamefont{Murphy}},
  \bibinfo{author}{\bibfnamefont{G.}~\bibnamefont{Cescutti}}, \bibnamefont{and}
  \bibinfo{author}{\bibfnamefont{C.}~\bibnamefont{Travaglio}},
  \bibinfo{journal}{Monthly Notices of the Royal Astronomical Society}
  \textbf{\bibinfo{volume}{474}}, \bibinfo{pages}{3133} (\bibinfo{year}{2018}).

\bibitem[{\citenamefont{Schatz et~al.}(1998)\citenamefont{Schatz, Aprahamian,
  G\"orres, Wiescher, Rauscher, Rembges, Thielemann, Pfeiffer, M\"oller, Kratz
  et~al.}}]{schatz1998}
\bibinfo{author}{\bibfnamefont{H.}~\bibnamefont{Schatz}},
  \bibinfo{author}{\bibfnamefont{A.}~\bibnamefont{Aprahamian}},
  \bibinfo{author}{\bibfnamefont{J.}~\bibnamefont{G\"orres}},
  \bibinfo{author}{\bibfnamefont{M.}~\bibnamefont{Wiescher}},
  \bibinfo{author}{\bibfnamefont{T.}~\bibnamefont{Rauscher}},
  \bibinfo{author}{\bibfnamefont{J.}~\bibnamefont{Rembges}},
  \bibinfo{author}{\bibfnamefont{F.-K.} \bibnamefont{Thielemann}},
  \bibinfo{author}{\bibfnamefont{B.}~\bibnamefont{Pfeiffer}},
  \bibinfo{author}{\bibfnamefont{P.}~\bibnamefont{M\"oller}},
  \bibinfo{author}{\bibfnamefont{K.-L.} \bibnamefont{Kratz}},
  \bibnamefont{et~al.}, \bibinfo{journal}{Physics Reports}
  \textbf{\bibinfo{volume}{294}}, \bibinfo{pages}{167} (\bibinfo{year}{1998}).

\bibitem[{\citenamefont{Sz\"ucs et~al.}(2014)\citenamefont{Sz\"ucs, Dillmann,
  Plag, and F\"ul\"op}}]{kadonis2014}
\bibinfo{author}{\bibfnamefont{T.}~\bibnamefont{Sz\"ucs}},
  \bibinfo{author}{\bibfnamefont{I.}~\bibnamefont{Dillmann}},
  \bibinfo{author}{\bibfnamefont{R.}~\bibnamefont{Plag}}, \bibnamefont{and}
  \bibinfo{author}{\bibfnamefont{Z.}~\bibnamefont{F\"ul\"op}},
  \bibinfo{journal}{Nuclear Data Sheets} \textbf{\bibinfo{volume}{120}},
  \bibinfo{pages}{191} (\bibinfo{year}{2014}),
  \bibinfo{note}{\mbox{\url{http://www.kadonis.org}}}.

\bibitem[{\citenamefont{Delbar et~al.}(1993)\citenamefont{Delbar, Galster,
  Leleux, Licot, Li\'{e}nard, Lipnik, Loiselet, Michotte, Ryckewaert, Vervier
  et~al.}}]{delbar1993}
\bibinfo{author}{\bibfnamefont{T.}~\bibnamefont{Delbar}},
  \bibinfo{author}{\bibfnamefont{W.}~\bibnamefont{Galster}},
  \bibinfo{author}{\bibfnamefont{P.}~\bibnamefont{Leleux}},
  \bibinfo{author}{\bibfnamefont{I.}~\bibnamefont{Licot}},
  \bibinfo{author}{\bibfnamefont{E.}~\bibnamefont{Li\'{e}nard}},
  \bibinfo{author}{\bibfnamefont{P.}~\bibnamefont{Lipnik}},
  \bibinfo{author}{\bibfnamefont{M.}~\bibnamefont{Loiselet}},
  \bibinfo{author}{\bibfnamefont{C.}~\bibnamefont{Michotte}},
  \bibinfo{author}{\bibfnamefont{G.}~\bibnamefont{Ryckewaert}},
  \bibinfo{author}{\bibfnamefont{J.}~\bibnamefont{Vervier}},
  \bibnamefont{et~al.}, \bibinfo{journal}{Physical Review C}
  \textbf{\bibinfo{volume}{48}}, \bibinfo{pages}{3088} (\bibinfo{year}{1993}).

\bibitem[{\citenamefont{D'Auria et~al.}(2004)\citenamefont{D'Auria, Azuma,
  Bishop, Buchmann, Chatterjee, Chen, Engel, Gigliotti, Greife, Hunter
  et~al.}}]{dauria2004}
\bibinfo{author}{\bibfnamefont{J.~M.} \bibnamefont{D'Auria}},
  \bibinfo{author}{\bibfnamefont{R.~E.} \bibnamefont{Azuma}},
  \bibinfo{author}{\bibfnamefont{S.}~\bibnamefont{Bishop}},
  \bibinfo{author}{\bibfnamefont{L.}~\bibnamefont{Buchmann}},
  \bibinfo{author}{\bibfnamefont{M.~L.} \bibnamefont{Chatterjee}},
  \bibinfo{author}{\bibfnamefont{A.~A.} \bibnamefont{Chen}},
  \bibinfo{author}{\bibfnamefont{S.}~\bibnamefont{Engel}},
  \bibinfo{author}{\bibfnamefont{D.}~\bibnamefont{Gigliotti}},
  \bibinfo{author}{\bibfnamefont{U.}~\bibnamefont{Greife}},
  \bibinfo{author}{\bibfnamefont{D.}~\bibnamefont{Hunter}},
  \bibnamefont{et~al.}, \bibinfo{journal}{Physical Review C}
  \textbf{\bibinfo{volume}{69}}, \bibinfo{pages}{065803} (\bibinfo{year}{2004}).

\bibitem[{\citenamefont{Akers et~al.}(2013)\citenamefont{Akers, Laird, Fulton,
  Ruiz, Bardayan, Buchmann, Christian, Davids, Erikson, Fallis
  et~al.}}]{akers2013}
\bibinfo{author}{\bibfnamefont{C.}~\bibnamefont{Akers}},
  \bibinfo{author}{\bibfnamefont{A.~M.} \bibnamefont{Laird}},
  \bibinfo{author}{\bibfnamefont{B.~R.} \bibnamefont{Fulton}},
  \bibinfo{author}{\bibfnamefont{C.}~\bibnamefont{Ruiz}},
  \bibinfo{author}{\bibfnamefont{D.~W.} \bibnamefont{Bardayan}},
  \bibinfo{author}{\bibfnamefont{L.}~\bibnamefont{Buchmann}},
  \bibinfo{author}{\bibfnamefont{G.}~\bibnamefont{Christian}},
  \bibinfo{author}{\bibfnamefont{B.}~\bibnamefont{Davids}},
  \bibinfo{author}{\bibfnamefont{L.}~\bibnamefont{Erikson}},
  \bibinfo{author}{\bibfnamefont{J.}~\bibnamefont{Fallis}},
  \bibnamefont{et~al.}, \bibinfo{journal}{Physical Review Letters}
  \textbf{\bibinfo{volume}{110}}, \bibinfo{pages}{262502} (\bibinfo{year}{2013}).

\bibitem[{\citenamefont{Lotay et~al.}(2016)\citenamefont{Lotay, Christian,
  Ruiz, Akers, Burke, Catford, Chen, Connolly, Davids, Fallis
  et~al.}}]{lotay2016}
\bibinfo{author}{\bibfnamefont{G.}~\bibnamefont{Lotay}},
  \bibinfo{author}{\bibfnamefont{G.}~\bibnamefont{Christian}},
  \bibinfo{author}{\bibfnamefont{C.}~\bibnamefont{Ruiz}},
  \bibinfo{author}{\bibfnamefont{C.}~\bibnamefont{Akers}},
  \bibinfo{author}{\bibfnamefont{D.}~\bibnamefont{Burke}},
  \bibinfo{author}{\bibfnamefont{W.}~\bibnamefont{Catford}},
  \bibinfo{author}{\bibfnamefont{A.}~\bibnamefont{Chen}},
  \bibinfo{author}{\bibfnamefont{D.}~\bibnamefont{Connolly}},
  \bibinfo{author}{\bibfnamefont{B.}~\bibnamefont{Davids}},
  \bibinfo{author}{\bibfnamefont{J.}~\bibnamefont{Fallis}},
  \bibnamefont{et~al.}, \bibinfo{journal}{Physical Review Letters}
  \textbf{\bibinfo{volume}{116}}, \bibinfo{pages}{132701} (\bibinfo{year}{2016}).

\bibitem[{gsi()}]{gsi}
\bibinfo{note}{{GSI: GSI Helmholtzzentrum f\"ur Schwerionenforschung},
  \url{http://www.gsi.de}}.

\bibitem[{fai()}]{fair}
\bibinfo{note}{{FAIR: Facility for Antiproton and Ion Research},
  \url{http://www.fair-center.eu}}.

\bibitem[{\citenamefont{Franzke}(1987)}]{franzke1987}
\bibinfo{author}{\bibfnamefont{B.}~\bibnamefont{Franzke}},
  \bibinfo{journal}{Nuclear Instruments and Methods in Physics Research Section
  B: Beam Interactions with Materials and Atoms}
  \textbf{\bibinfo{volume}{24-25}}, \bibinfo{pages}{18 }
  (\bibinfo{year}{1987}).

\bibitem[{\citenamefont{Geissel et~al.}(1992)\citenamefont{Geissel, Armbruster,
  Behr, Br\"unle, Burkard, Chen, Folger, Franczak, Keller, Klepper
  et~al.}}]{geissel1992}
\bibinfo{author}{\bibfnamefont{H.}~\bibnamefont{Geissel}},
  \bibinfo{author}{\bibfnamefont{P.}~\bibnamefont{Armbruster}},
  \bibinfo{author}{\bibfnamefont{K.}~\bibnamefont{Behr}},
  \bibinfo{author}{\bibfnamefont{A.}~\bibnamefont{Br\"unle}},
  \bibinfo{author}{\bibfnamefont{K.}~\bibnamefont{Burkard}},
  \bibinfo{author}{\bibfnamefont{M.}~\bibnamefont{Chen}},
  \bibinfo{author}{\bibfnamefont{H.}~\bibnamefont{Folger}},
  \bibinfo{author}{\bibfnamefont{B.}~\bibnamefont{Franczak}},
  \bibinfo{author}{\bibfnamefont{H.}~\bibnamefont{Keller}},
  \bibinfo{author}{\bibfnamefont{O.}~\bibnamefont{Klepper}},
  \bibnamefont{et~al.}, \bibinfo{journal}{Nuclear Instruments and Methods in
  Physics Research Section B: Beam Interactions with Materials and Atoms}
  \textbf{\bibinfo{volume}{70}}, \bibinfo{pages}{286} (\bibinfo{year}{1992}).

\bibitem[{\citenamefont{Mei et~al.}(2015)\citenamefont{Mei, Aumann, Bishop,
  Blaum, Boretzky, Bosch, Brandau, Br\"auning, Davinson, Dillmann
  et~al.}}]{mei2015}
\bibinfo{author}{\bibfnamefont{B.}~\bibnamefont{Mei}},
  \bibinfo{author}{\bibfnamefont{T.}~\bibnamefont{Aumann}},
  \bibinfo{author}{\bibfnamefont{S.}~\bibnamefont{Bishop}},
  \bibinfo{author}{\bibfnamefont{K.}~\bibnamefont{Blaum}},
  \bibinfo{author}{\bibfnamefont{K.}~\bibnamefont{Boretzky}},
  \bibinfo{author}{\bibfnamefont{F.}~\bibnamefont{Bosch}},
  \bibinfo{author}{\bibfnamefont{C.}~\bibnamefont{Brandau}},
  \bibinfo{author}{\bibfnamefont{H.}~\bibnamefont{Br\"auning}},
  \bibinfo{author}{\bibfnamefont{T.}~\bibnamefont{Davinson}},
  \bibinfo{author}{\bibfnamefont{I.}~\bibnamefont{Dillmann}},
  \bibnamefont{et~al.}, \bibinfo{journal}{Physical Review C}
  \textbf{\bibinfo{volume}{92}}, \bibinfo{pages}{035803} (\bibinfo{year}{2015}).

\bibitem[{\citenamefont{Litvinov et~al.}(2013)\citenamefont{Litvinov, Bishop,
  Blaum, Bosch, Brandau, Chen, Dillmann, Egelhof, Geissel, Grisenti
  et~al.}}]{litvinov2013}
\bibinfo{author}{\bibfnamefont{Y.}~\bibnamefont{Litvinov}},
  \bibinfo{author}{\bibfnamefont{S.}~\bibnamefont{Bishop}},
  \bibinfo{author}{\bibfnamefont{K.}~\bibnamefont{Blaum}},
  \bibinfo{author}{\bibfnamefont{F.}~\bibnamefont{Bosch}},
  \bibinfo{author}{\bibfnamefont{C.}~\bibnamefont{Brandau}},
  \bibinfo{author}{\bibfnamefont{L.}~\bibnamefont{Chen}},
  \bibinfo{author}{\bibfnamefont{I.}~\bibnamefont{Dillmann}},
  \bibinfo{author}{\bibfnamefont{P.}~\bibnamefont{Egelhof}},
  \bibinfo{author}{\bibfnamefont{H.}~\bibnamefont{Geissel}},
  \bibinfo{author}{\bibfnamefont{R.}~\bibnamefont{Grisenti}},
  \bibnamefont{et~al.}, \bibinfo{journal}{Nuclear Instruments and Methods in
  Physics Research Section B: Beam Interactions with Materials and Atoms}
  \textbf{\bibinfo{volume}{317}}, \bibinfo{pages}{603} (\bibinfo{year}{2013}).

\bibitem[{\citenamefont{K\"uhnel et~al.}(2009)\citenamefont{K\"uhnel, Petridis,
  Winters, Popp, D\"orner, St\"ohlker, and Grisenti}}]{target}
\bibinfo{author}{\bibfnamefont{M.}~\bibnamefont{K\"uhnel}},
  \bibinfo{author}{\bibfnamefont{N.}~\bibnamefont{Petridis}},
  \bibinfo{author}{\bibfnamefont{D.}~\bibnamefont{Winters}},
  \bibinfo{author}{\bibfnamefont{U.}~\bibnamefont{Popp}},
  \bibinfo{author}{\bibfnamefont{R.}~\bibnamefont{D\"orner}},
  \bibinfo{author}{\bibfnamefont{T.}~\bibnamefont{St\"ohlker}},
  \bibnamefont{and} \bibinfo{author}{\bibfnamefont{R.}~\bibnamefont{Grisenti}},
  \bibinfo{journal}{Nuclear Instruments and Methods in Physics Research Section
  A: Accelerators, Spectrometers, Detectors and Associated Equipment}
  \textbf{\bibinfo{volume}{602}}, \bibinfo{pages}{311} (\bibinfo{year}{2009}).

\bibitem[{\citenamefont{Ruiz et~al.}(2014)\citenamefont{Ruiz, Greife, and
  Hager}}]{ruiz2014}
\bibinfo{author}{\bibfnamefont{C.}~\bibnamefont{Ruiz}},
  \bibinfo{author}{\bibfnamefont{U.}~\bibnamefont{Greife}}, \bibnamefont{and}
  \bibinfo{author}{\bibfnamefont{U.}~\bibnamefont{Hager}},
  \bibinfo{journal}{The European Physical Journal A}
  \textbf{\bibinfo{volume}{50}}, \bibinfo{pages}{99} (\bibinfo{year}{2014}).

\bibitem[{\citenamefont{Grieser et~al.}(2012)\citenamefont{Grieser, Litvinov,
  Raabe, Blaum, Blumenfeld, Butler, Wenander, Woods, Aliotta, Andreyev
  et~al.}}]{grieser2012}
\bibinfo{author}{\bibfnamefont{M.}~\bibnamefont{Grieser}},
  \bibinfo{author}{\bibfnamefont{Y.~A.} \bibnamefont{Litvinov}},
  \bibinfo{author}{\bibfnamefont{R.}~\bibnamefont{Raabe}},
  \bibinfo{author}{\bibfnamefont{K.}~\bibnamefont{Blaum}},
  \bibinfo{author}{\bibfnamefont{Y.}~\bibnamefont{Blumenfeld}},
  \bibinfo{author}{\bibfnamefont{P.~A.} \bibnamefont{Butler}},
  \bibinfo{author}{\bibfnamefont{F.}~\bibnamefont{Wenander}},
  \bibinfo{author}{\bibfnamefont{P.~J.} \bibnamefont{Woods}},
  \bibinfo{author}{\bibfnamefont{M.}~\bibnamefont{Aliotta}},
  \bibinfo{author}{\bibfnamefont{A.}~\bibnamefont{Andreyev}},
  \bibnamefont{et~al.}, \bibinfo{journal}{The European Physical Journal Special
  Topics} \textbf{\bibinfo{volume}{207}}, \bibinfo{pages}{1}
  (\bibinfo{year}{2012}).

\bibitem[{\citenamefont{St\"ohlker et~al.}(1998)\citenamefont{St\"ohlker,
  Ludziejewski, Reich, Bosch, Dunford, Eichler, Franzke, Kozhuharov, Menzel
  et~al.}}]{stoehlker1998}
\bibinfo{author}{\bibfnamefont{T.}~\bibnamefont{St\"ohlker}},
  \bibinfo{author}{\bibfnamefont{T.}~\bibnamefont{Ludziejewski}},
  \bibinfo{author}{\bibfnamefont{H.}~\bibnamefont{Reich}},
  \bibinfo{author}{\bibfnamefont{F.}~\bibnamefont{Bosch}},
  \bibinfo{author}{\bibfnamefont{R.~W.} \bibnamefont{Dunford}},
  \bibinfo{author}{\bibfnamefont{J.}~\bibnamefont{Eichler}},
  \bibinfo{author}{\bibfnamefont{B.}~\bibnamefont{Franzke}},
  \bibinfo{author}{\bibfnamefont{C.}~\bibnamefont{Kozhuharov}},
  \bibinfo{author}{\bibfnamefont{G.}~\bibnamefont{Menzel}},
  \bibnamefont{et~al.}, \bibinfo{journal}{Physical Review A}
  \textbf{\bibinfo{volume}{58}}, \bibinfo{pages}{2043} (\bibinfo{year}{1998}).

\bibitem[{\citenamefont{Klepper and Kozhuharov}(2003)}]{klepper2003}
\bibinfo{author}{\bibfnamefont{O.}~\bibnamefont{Klepper}} \bibnamefont{and}
  \bibinfo{author}{\bibfnamefont{C.}~\bibnamefont{Kozhuharov}},
  \bibinfo{journal}{Nuclear Instruments and Methods in Physics Research Section
  B: Beam Interactions with Materials and Atoms}
  \textbf{\bibinfo{volume}{204}}, \bibinfo{pages}{553} (\bibinfo{year}{2003}).

\bibitem[{dss()}]{dsssd}
\bibinfo{note}{{Micron Semiconductor LTD},
  \url{http://www.micronsemiconductor.co.uk}}.

\bibitem[{\citenamefont{Eichler and St\"ohlker}(2007)}]{eichler2007}
\bibinfo{author}{\bibfnamefont{J.}~\bibnamefont{Eichler}} \bibnamefont{and}
  \bibinfo{author}{\bibfnamefont{T.}~\bibnamefont{St\"ohlker}},
  \bibinfo{journal}{Physics Reports} \textbf{\bibinfo{volume}{439}},
  \bibinfo{pages}{1} (\bibinfo{year}{2007}).

\bibitem[{\citenamefont{Artemyev et~al.}(2010)\citenamefont{Artemyev,
  Surzhykov, Fritzsche, Najjari, and Voitkiv}}]{artemyev2010}
\bibinfo{author}{\bibfnamefont{A.~N.} \bibnamefont{Artemyev}},
  \bibinfo{author}{\bibfnamefont{A.}~\bibnamefont{Surzhykov}},
  \bibinfo{author}{\bibfnamefont{S.}~\bibnamefont{Fritzsche}},
  \bibinfo{author}{\bibfnamefont{B.}~\bibnamefont{Najjari}}, \bibnamefont{and}
  \bibinfo{author}{\bibfnamefont{A.~B.} \bibnamefont{Voitkiv}},
  \bibinfo{journal}{Phys. Rev. A} \textbf{\bibinfo{volume}{82}},
  \bibinfo{pages}{022716} (\bibinfo{year}{2010}).

\bibitem[{\citenamefont{Yorkston et~al.}(1987)\citenamefont{Yorkston, Shotter,
  Syme, and Huxtable}}]{yorkston1987}
\bibinfo{author}{\bibfnamefont{J.}~\bibnamefont{Yorkston}},
  \bibinfo{author}{\bibfnamefont{A.}~\bibnamefont{Shotter}},
  \bibinfo{author}{\bibfnamefont{D.}~\bibnamefont{Syme}}, \bibnamefont{and}
  \bibinfo{author}{\bibfnamefont{G.}~\bibnamefont{Huxtable}},
  \bibinfo{journal}{Nuclear Instruments and Methods in Physics Research Section
  A: Accelerators, Spectrometers, Detectors and Associated Equipment}
  \textbf{\bibinfo{volume}{262}}, \bibinfo{pages}{353} (\bibinfo{year}{1987}).

\bibitem[{\citenamefont{Iwasa et~al.}(2011)\citenamefont{Iwasa, Weick, and
  Geissel}}]{iwasa2011}
\bibinfo{author}{\bibfnamefont{N.}~\bibnamefont{Iwasa}},
  \bibinfo{author}{\bibfnamefont{H.}~\bibnamefont{Weick}}, \bibnamefont{and}
  \bibinfo{author}{\bibfnamefont{H.}~\bibnamefont{Geissel}},
  \bibinfo{journal}{Nuclear Instruments and Methods in Physics Research Section
  B: Beam Interactions with Materials and Atoms}
  \textbf{\bibinfo{volume}{269}}, \bibinfo{pages}{752} (\bibinfo{year}{2011}).

\bibitem[{\citenamefont{Rutherford}(1911)}]{rutherford1911}
\bibinfo{author}{\bibfnamefont{E.}~\bibnamefont{Rutherford}},
  \bibinfo{journal}{The London, Edinburgh, and Dublin Philosophical Magazine
  and Journal of Science} \textbf{\bibinfo{volume}{21}}, \bibinfo{pages}{669}
  (\bibinfo{year}{1911}).
  
  \bibitem[{\citenamefont{Nurmela et~al.}(1998)\citenamefont{Nurmela, Zazubovich,
  R\"ais\"anen, Rauhala, and Lappalainen}}]{nurmela1998}
\bibinfo{author}{\bibfnamefont{A.}~\bibnamefont{Nurmela}},
  \bibinfo{author}{\bibfnamefont{V.}~\bibnamefont{Zazubovich}},
  \bibinfo{author}{\bibfnamefont{J.}~\bibnamefont{R\"ais\"anen}},
  \bibinfo{author}{\bibfnamefont{E.}~\bibnamefont{Rauhala}}, \bibnamefont{and}
  \bibinfo{author}{\bibfnamefont{R.}~\bibnamefont{Lappalainen}},
  \bibinfo{journal}{Journal of Applied Physics} \textbf{\bibinfo{volume}{84}},
  \bibinfo{pages}{1796} (\bibinfo{year}{1998}).

\bibitem[{\citenamefont{Ullmann et~al.}(2015)\citenamefont{Ullmann, Andelkovic,
  Dax, Geithner, Geppert, Gorges, Hammen, Hannen, Kaufmann, K\"onig
  et~al.}}]{ullmann2015}
\bibinfo{author}{\bibfnamefont{J.}~\bibnamefont{Ullmann}},
  \bibinfo{author}{\bibfnamefont{Z.}~\bibnamefont{Andelkovic}},
  \bibinfo{author}{\bibfnamefont{A.}~\bibnamefont{Dax}},
  \bibinfo{author}{\bibfnamefont{W.}~\bibnamefont{Geithner}},
  \bibinfo{author}{\bibfnamefont{C.}~\bibnamefont{Geppert}},
  \bibinfo{author}{\bibfnamefont{C.}~\bibnamefont{Gorges}},
  \bibinfo{author}{\bibfnamefont{M.}~\bibnamefont{Hammen}},
  \bibinfo{author}{\bibfnamefont{V.}~\bibnamefont{Hannen}},
  \bibinfo{author}{\bibfnamefont{S.}~\bibnamefont{Kaufmann}},
  \bibinfo{author}{\bibfnamefont{K.}~\bibnamefont{K\"onig}},
  \bibnamefont{et~al.}, \bibinfo{journal}{Journal of Physics B: Atomic,
  Molecular and Optical Physics} \textbf{\bibinfo{volume}{48}},
  \bibinfo{pages}{144022} (\bibinfo{year}{2015}).

\bibitem[{\citenamefont{Hauser and Feshbach}(1952)}]{hauserfeshbach}
\bibinfo{author}{\bibfnamefont{W.}~\bibnamefont{Hauser}} \bibnamefont{and}
  \bibinfo{author}{\bibfnamefont{H.}~\bibnamefont{Feshbach}},
  \bibinfo{journal}{Physical Review} \textbf{\bibinfo{volume}{87}},
  \bibinfo{pages}{366} (\bibinfo{year}{1952}).

\bibitem[{\citenamefont{Rauscher}(2011)}]{rauscher11}
\bibinfo{author}{\bibfnamefont{T.}~\bibnamefont{Rauscher}},
  \bibinfo{journal}{International Journal of Modern Physics E}
  \textbf{\bibinfo{volume}{20}}, \bibinfo{pages}{1071} (\bibinfo{year}{2011}).

\bibitem[{\citenamefont{Rauscher}()}]{smaragd}
\bibinfo{author}{\bibfnamefont{T.}~\bibnamefont{Rauscher}}, \bibinfo{note}{code
  SMARAGD, version 0.10.0s}.

\bibitem[{\citenamefont{Rauscher and Thielemann}(2000)}]{rauscher00}
\bibinfo{author}{\bibfnamefont{T.}~\bibnamefont{Rauscher}} \bibnamefont{and}
  \bibinfo{author}{\bibfnamefont{F.-K.} \bibnamefont{Thielemann}},
  \bibinfo{journal}{Atomic Data and Nuclear Data Tables}
  \textbf{\bibinfo{volume}{75}}, \bibinfo{pages}{1} (\bibinfo{year}{2000}).

\bibitem[{\citenamefont{Cyburt et~al.}(2010)\citenamefont{Cyburt, Amthor,
  Ferguson, Meisel, Smith, Warren, Heger, Hoffman, Rauscher, Sakharuk
  et~al.}}]{cyburt2010}
\bibinfo{author}{\bibfnamefont{R.~H.} \bibnamefont{Cyburt}},
  \bibinfo{author}{\bibfnamefont{A.~M.} \bibnamefont{Amthor}},
  \bibinfo{author}{\bibfnamefont{R.}~\bibnamefont{Ferguson}},
  \bibinfo{author}{\bibfnamefont{Z.}~\bibnamefont{Meisel}},
  \bibinfo{author}{\bibfnamefont{K.}~\bibnamefont{Smith}},
  \bibinfo{author}{\bibfnamefont{S.}~\bibnamefont{Warren}},
  \bibinfo{author}{\bibfnamefont{A.}~\bibnamefont{Heger}},
  \bibinfo{author}{\bibfnamefont{R.~D.} \bibnamefont{Hoffman}},
  \bibinfo{author}{\bibfnamefont{T.}~\bibnamefont{Rauscher}},
  \bibinfo{author}{\bibfnamefont{A.}~\bibnamefont{Sakharuk}},
  \bibnamefont{et~al.}, \bibinfo{journal}{The Astrophysical Journal Supplement}
  \textbf{\bibinfo{volume}{189}}, \bibinfo{pages}{240} (\bibinfo{year}{2010}).

\bibitem[{\citenamefont{Jeukenne et~al.}(1977)\citenamefont{Jeukenne, Lejeune,
  and Mahaux}}]{jeukenne77}
\bibinfo{author}{\bibfnamefont{J.~P.} \bibnamefont{Jeukenne}},
  \bibinfo{author}{\bibfnamefont{A.}~\bibnamefont{Lejeune}}, \bibnamefont{and}
  \bibinfo{author}{\bibfnamefont{C.}~\bibnamefont{Mahaux}},
  \bibinfo{journal}{Physical Review C} \textbf{\bibinfo{volume}{16}},
  \bibinfo{pages}{80} (\bibinfo{year}{1977}).

\bibitem[{\citenamefont{Lejeune}(1980)}]{lejeune80}
\bibinfo{author}{\bibfnamefont{A.}~\bibnamefont{Lejeune}},
  \bibinfo{journal}{Physical Review C} \textbf{\bibinfo{volume}{21}},
  \bibinfo{pages}{1107} (\bibinfo{year}{1980}).

\bibitem[{\citenamefont{Rauscher et~al.}(1997)\citenamefont{Rauscher,
  Thielemann, and Kratz}}]{rtk}
\bibinfo{author}{\bibfnamefont{T.}~\bibnamefont{Rauscher}},
  \bibinfo{author}{\bibfnamefont{F.-K.} \bibnamefont{Thielemann}},
  \bibnamefont{and} \bibinfo{author}{\bibfnamefont{K.-L.} \bibnamefont{Kratz}},
  \bibinfo{journal}{Physical Review C} \textbf{\bibinfo{volume}{56}},
  \bibinfo{pages}{1613} (\bibinfo{year}{1997}).

\bibitem[{\citenamefont{Mocelj et~al.}(2007)\citenamefont{Mocelj, Rauscher,
  Mart\'inez-Pinedo, Langanke, Pacearescu, F\"a{\ss}ler, Thielemann, and
  Alhassid}}]{mocelj}
\bibinfo{author}{\bibfnamefont{D.}~\bibnamefont{Mocelj}},
  \bibinfo{author}{\bibfnamefont{T.}~\bibnamefont{Rauscher}},
  \bibinfo{author}{\bibfnamefont{G.}~\bibnamefont{Mart\'inez-Pinedo}},
  \bibinfo{author}{\bibfnamefont{K.}~\bibnamefont{Langanke}},
  \bibinfo{author}{\bibfnamefont{L.}~\bibnamefont{Pacearescu}},
  \bibinfo{author}{\bibfnamefont{A.}~\bibnamefont{F\"a{\ss}ler}},
  \bibinfo{author}{\bibfnamefont{F.-K.} \bibnamefont{Thielemann}},
  \bibnamefont{and} \bibinfo{author}{\bibfnamefont{Y.}~\bibnamefont{Alhassid}},
  \bibinfo{journal}{Physical Review C} \textbf{\bibinfo{volume}{75}},
  \bibinfo{pages}{045807} (\bibinfo{year}{2007}).

\bibitem[{\citenamefont{Rauscher}(2012)}]{rauscher12}
\bibinfo{author}{\bibfnamefont{T.}~\bibnamefont{Rauscher}},
  \bibinfo{journal}{Astrophysical Journal Supplement}
  \textbf{\bibinfo{volume}{201}}, \bibinfo{pages}{26} (\bibinfo{year}{2012}).

\bibitem[{\citenamefont{Cyburt et~al.}(2016)\citenamefont{Cyburt, Amthor,
  Heger, Johnson, Keek, Meisel, Schatz, and Smith}}]{cyburt2016}
\bibinfo{author}{\bibfnamefont{R.~H.} \bibnamefont{Cyburt}},
  \bibinfo{author}{\bibfnamefont{A.~M.} \bibnamefont{Amthor}},
  \bibinfo{author}{\bibfnamefont{A.}~\bibnamefont{Heger}},
  \bibinfo{author}{\bibfnamefont{E.}~\bibnamefont{Johnson}},
  \bibinfo{author}{\bibfnamefont{L.}~\bibnamefont{Keek}},
  \bibinfo{author}{\bibfnamefont{Z.}~\bibnamefont{Meisel}},
  \bibinfo{author}{\bibfnamefont{H.}~\bibnamefont{Schatz}}, \bibnamefont{and}
  \bibinfo{author}{\bibfnamefont{K.}~\bibnamefont{Smith}},
  \bibinfo{journal}{The Astrophysical Journal} \textbf{\bibinfo{volume}{830}},
  \bibinfo{pages}{55} (\bibinfo{year}{2016}).

\bibitem[{\citenamefont{Lestinsky et~al.}(2016)\citenamefont{Lestinsky,
  Andrianov, Aurand, Bagnoud, Bernhardt, Beyer, Bishop, Blaum, Bleile, Borovik
  et~al.}}]{lestinsky2016}
\bibinfo{author}{\bibfnamefont{M.}~\bibnamefont{Lestinsky}},
  \bibinfo{author}{\bibfnamefont{V.}~\bibnamefont{Andrianov}},
  \bibinfo{author}{\bibfnamefont{B.}~\bibnamefont{Aurand}},
  \bibinfo{author}{\bibfnamefont{V.}~\bibnamefont{Bagnoud}},
  \bibinfo{author}{\bibfnamefont{D.}~\bibnamefont{Bernhardt}},
  \bibinfo{author}{\bibfnamefont{H.}~\bibnamefont{Beyer}},
  \bibinfo{author}{\bibfnamefont{S.}~\bibnamefont{Bishop}},
  \bibinfo{author}{\bibfnamefont{K.}~\bibnamefont{Blaum}},
  \bibinfo{author}{\bibfnamefont{A.}~\bibnamefont{Bleile}},
  \bibinfo{author}{\bibfnamefont{A.}~\bibnamefont{Borovik}},
  \bibnamefont{et~al.}, \bibinfo{journal}{The European Physical Journal Special
  Topics} \textbf{\bibinfo{volume}{225}}, \bibinfo{pages}{797}
  (\bibinfo{year}{2016}).

\end{thebibliography}
\end{document}